# Uncovering and experimental realization of multimodal 3D topological metamaterials for low-frequency and multiband elastic wave control

Patrick Dorin, Mustafa Khan, K. W. Wang

Department of Mechanical Engineering, University of Michigan, Ann Arbor, MI 48109, USA

**Email:** pdorin@umich.edu

**Abstract**

Topological metamaterials unlock confined and robust elastic wave control in mechanical structures. Recent breakthroughs have precipitated the development of 3D topological mechanical metamaterials, which extend beyond the conventional 1D and 2D metamaterials to facilitate extraordinary wave manipulation along 2D planar and layer-dependent elastic waveguides. While promising, significant research gaps exist that impede the practical implementation of 3D topological metamaterials. The 3D topological metamaterials studied thus far are constrained to function in single frequency bandwidths that are typically in a high-frequency regime, and a comprehensive experimental investigation remains elusive. In this paper, we address these research gaps and advance the state of the art through the synthesis and experimental realization of a 3D topological metamaterial that exploits multimodal local resonance to enable low-frequency elastic wave control over multiple distinct frequency bands. The proposed metamaterial is geometrically configured to create multimodal local resonators whose frequency characteristics govern the emergence of four unique low-frequency topological states. Numerical simulations uncover how these topological states can be employed to achieve polarization-, frequency-, and layer-dependent wave manipulation in 3D structures. An experimental study results in the attainment of complete wave fields that unambiguously illustrate 2D topological waveguides and multi-polarized wave control in a physical testbed. The outcomes from this work open the door for future research with 3D topological mechanical metamaterials and reveal the applicability of the proposed metamaterial for various wave control applications.

**Significance Statement**

3D topological metamaterials have advanced beyond 1D and 2D systems to enable unprecedented control over the flow of elastic waves in mechanical structures. However, the 3D topological mechanical metamaterials studied thus far are largely limited to operating in single- and high-frequency regimes, and a comprehensive experimental study remains elusive. In this paper, we synthesize a 3D topological metamaterial that harnesses multimodal local resonance to unlock multiband and low-frequency elastic wave control. This work leverages innovative fabrication and measurement techniques to produce unequivocal experimental evidence of 2D planar and multi-polarized elastic waveguides. The outcomes from this investigation will pave the way for future research and practical engineering implementations of 3D wave-based mechanical devices that process information, isolate vibrations, and harvest energy.

**Main Text**

**Introduction**
Topological phases have been employed to achieve robust electron transport in quantum systems through conducting states that are protected from local perturbations [1]–[4]. Recently, topological phases have been integrated into the synthesis of elastic (i.e., mechanical) metamaterials, facilitating extraordinary control over the flow of energy and information contained by elastic waves in mechanical systems. These so-called topological metamaterials enable low-loss transport and arbitrary directional manipulation of elastic waves via localized topological states that are protected from unwanted scattering in the presence of structural defects or disorder [5]–[9]. The remarkable capabilities and robustness of topological metamaterials have been exploited to enhance performance in technical applications that include vibration energy harvesters [10]–[13], on-chip communications [14]–[17], elastic antennas [18], isolators [19], and mechanical information processing [20]–[22].

Initial research concerning topological metamaterials focused on the theoretical prediction and experimental demonstration of zero-dimensional (0D) topological states in one-dimensional (1D) mechanical structures (e.g., the wave is localized to a point in a rod) and 1D topological states in two-dimensional (2D) mechanical structures (e.g., the wave is localized along a line waveguide in a thin plate) [23]–[35]. Building upon the promising initial outcomes, researchers have begun to explore beyond the traditional 1D and 2D systems to achieve 2D topological states in three-dimensional (3D) structures (e.g., the wave is localized along a planar waveguide in a 3D cubic geometry). To construct 3D topological metamaterials, the elastic analogs of Weyl semimetals or the quantum valley Hall effect (QVHE) from electronic systems have been created by carefully configuring the spatial symmetries of 3D periodic lattice geometries [36]–[42]. The previous research on 3D topological metamaterials has uncovered multiple unprecedented functionalities, including robust elastic wave manipulation along 2D planar waveguides in numerous spatial directions, multifaceted wave splitters/networks, and layer-selective wave control.

While 3D topological mechanical metamaterials have exhibited the potential to surpass what is possible in 1D and 2D metamaterials, research gaps exist that inhibit their successful implementation in practice. Despite numerous theoretical studies [37]–[42], there is very little experimental evidence of elastic wave control in 3D topological metamaterials [36], due to the challenges associated with the fabrication and testing of intricate 3D mechanical architectures. Furthermore, despite previous studies demonstrating multiband operation in 2D topological metamaterials [43]–[48], the 3D topological metamaterials established thus far are constrained to function in a single frequency band. This single-band characteristic limits the working bandwidth and information-carrying capacity of 3D topological metamaterials, reducing their suitability for multiband wave-based applications such as lasers [49], filters [50], resonators [51], on-chip circuits [52], isolators [53], [54], and wireless networks [55]. Finally, due to a typical dependence on the Bragg scattering mechanism, the previously developed 3D topological metamaterials largely function in a high-frequency regime (i.e., the ultrasonic range: 20 kHz to ~1 GHz) and would need to be scaled to impractically large dimensions to achieve low-frequency (i.e., few Hz to 20 kHz) elastic wave manipulation.

This research addresses the aforementioned gaps and advances the state of the art through the synthesis and experimental realization of a novel 3D topological metamaterial that harnesses multimodal local resonance for low-frequency and multiband elastic wave control. The geometry of the proposed mechanical metamaterial is configured to obtain four distinct low-frequency (<6 kHz) topological states derived from multimodal local resonances that can be tailored without changing the lattice constant. Rich polarization-dependent behavior is encoded into the metamaterial by taking advantage of the multiple unique polarizations of the resonant topological modes. Dispersion analyses and full-scale response simulations illustrate how these topological states emerge through the QVHE and can be exploited to obtain frequency- and layer-dependent elastic waveguides in 3D structures. Moreover, a comprehensive experimental investigation validates the theoretical predictions and produces the first measurements of complete wave fields and multi-polarized elastic waveguides in a 3D topological metamaterial. The findings presented within this paper pave the way for future experimental research on 3D topological mechanical metamaterials and illuminate the multifaceted features of the proposed metamaterial that would be of great benefit for various applications.

## Results
### 3D Metamaterial Description
The proposed metamaterial is a 3D periodic structure with aluminum interconnecting rods (for interconnection in the $z$ direction, with radius $r_r$ = 0.93 mm) and two resonators in the unit cell (Figure 1a,b). The schematic for the unit cell is given in Figure 1b, where $a$ = 50 mm is the in-plane ($x$-$y$) lattice constant, $h_o$ = 25 mm is the out-of-plane ($z$) lattice constant, and $\vec{a}_1 = a\hat{\imath}$, $\vec{a}_2 = a(1/2\hat{\imath} + \sqrt{3}/2\hat{\jmath})$, and $\vec{a}_3 = h_o\hat{k}$ are the lattice basis vectors. The resonators are created using steel masses ($E_m$ = 200 GPa, $\rho_m$ = 8000 kg·m$^{-3}$, $\nu_m$ = 0.30) of radius $r_m$ = 7.0 mm and baseline height $h_m$ = 14.3 mm that are attached to aluminum spring ligaments ($E_l$ = 69 GPa, $\rho_l$ = 2700 kg·m$^{-3}$, $\nu_l$ = 0.33) of height $h_l$ = 1.5 mm and width $w_{l1,l2}$ = 1.5 mm. Each resonator mass is comprised of two sub-masses with heights defined as $h_{m1} = \frac{h_m}{2}(1-\alpha)$ and $h_{m2} = \frac{h_m}{2}(1+\alpha)$, where $\alpha$ is the mass height perturbation parameter. The local resonance phenomenon, which enables wave control near the natural frequencies of the resonators [56], is exploited to obtain multiband and low-frequency wave dispersion characteristics through the design of the resonant elements. The lattice is arranged into a honeycomb pattern with $D_{6h}$ lattice symmetry to acquire non-trivial topological features through the QVHE [57]–[60].

### Unit Cell Dispersion and Multimodal Resonance Effect
The wave dispersion characteristics of the unit cell are analyzed first to investigate the topological characteristics and unique wave propagation features of the proposed metamaterial. As shown in Figure 1c, a unit cell with equivalent mass heights ($\alpha$ = 0) is designated as a Type 0 lattice. The band structure is calculated using the commercial finite element solver COMSOL Multiphysics (see Section S1 of the Supporting Information for further details on the simulation methods). The $D_{6h}$ symmetry present in the Type 0 lattice results in four distinct Dirac nodal line degeneracies appearing in the band structure across a wide frequency range (Figure 1c) [1], [6], [59]. These Dirac degeneracies are labeled as D1, D2, D3, and D4 in order of lowest to highest frequency, $f_{d-D1}$ = 0.37 kHz, $f_{d-D2}$ = 1.41 kHz, $f_{d-D3}$ = 3.66 kHz, and $f_{d-D4}$ = 4.76 kHz, which is computed at the midpoint of the K-H line. A polarization parameter $\Pi = \frac{\iiint_{V_U}|w|^2 dV}{\iiint_{V_U}|u|^2+|v|^2+|w|^2 dV}$ is defined and measured for each mode in the band structure, where $V_U$ is the volume of the unit cell, and $u$, $v$, and $w$ are the displacement components in the $x$, $y$, and $z$ directions, respectively. The Dirac degeneracies in Figure 1c each contain different polarizations, with D1 being predominantly out-of-plane polarized ($\Pi \approx 1$, yellow in Figure 1c), D2 having a mixed polarization ($\Pi = 0.5$, green in Figure 1c), and D3/D4 containing in-plane polarized displacements ($\Pi \approx 0$, blue in Figure 1c). When the mass heights are perturbed from the baseline ($\alpha \neq 0$), the $D_{6h}$ symmetry is reduced to $D_{3h}$ symmetry, and the lattice is designated as Type A for $\alpha$ < 0 or Type B for $\alpha$ > 0. Calculation of the band structure for the cases of $\alpha$ = ±0.11 reveals that this reduction in symmetry splits the Dirac degeneracies and leads to bandgaps opening along the K-H line (Figure 1d, note that the Type A and Type B band structures are superimposed, as they are identical for $|\alpha|$ = 0.11). For the case of $|\alpha|$ = 0.11, partial or complete bandgaps are obtained from D2, D3, and D4, which are shaded in gray in Figure 1d and cover the frequency ranges of $f$ = 1.3 to 1.5 kHz, $f$ = 3.5 to 3.8 kHz, and $f$ = 4.5 to 5.1 kHz, respectively. A bandgap may also be opened from D1 if the mass perturbation $|\alpha|$ is increased further (see Section S2 of the Supporting Information). The topological nature of the D2, D3, and D4 bandgaps is evaluated by computing the valley Chern number $C_{v-p} = \frac{1}{2\pi}\iint_v B_p(k)d^2k$ for the bands bordering each bandgap, where $B_p(k)$ is the Berry curvature, $p$ = 1 refers to the band delineating the low-frequency bandgap boundary, and $p$ = 2 refers to the band delineating the high-frequency bandgap boundary (see Section S3 of the Supporting Information for more details on $C_{v-p}$ calculations). The resulting $C_{v-p}$ are $C_{v-1}^{Type\ A}$ = 0.11, $C_{v-2}^{Type\ A}$ = -0.09, $C_{v-1}^{Type\ B}$ = -0.11, $C_{v-2}^{Type\ B}$ = 0.09 for the D2 bandgap; $C_{v-1}^{Type\ A}$ = -0.14, $C_{v-2}^{Type\ A}$ = 0.17, $C_{v-1}^{Type\ B}$ = 0.15, $C_{v-2}^{Type\ B}$ = -0.18 for the D3 bandgap; and $C_{v-1}^{Type\ A}$ = 0.25, $C_{v-2}^{Type\ A}$ = -0.30, $C_{v-1}^{Type\ B}$ = -0.26, $C_{v-2}^{Type\ B}$ = 0.30 for the D4 bandgap. The nonzero $C_{v-p}$ values reveal the topological characteristic for each bandgap, while the equal and inverted $C_{v-p}$ calculated for the Type A and Type B configurations indicate that they are topologically distinct.

Further analysis of the unit cell dispersion uncovers the multimodal resonant characteristic of the proposed 3D metamaterial and elucidates how this feature enables low-frequency topological bandgaps over multiple frequency bands. The mode shapes for the bands bordering each of the four topological bandgaps in the $|\alpha|$=0.11 case are shown in Figure 1e. All of these mode shapes contain displacement that is largely confined to the resonator masses, a distinguishing characteristic of the local resonance mechanism for bandgap formation [56], [61]. Furthermore, each set of modes displays a distinct resonant behavior: out-of-plane translational for D1 ($\Pi$≈1), rocking for D2 ($\Pi$=0.6), in-plane twisting for D3 ($\Pi$≈0), and in-plane translational for D4 ($\Pi$≈0), illustrating the multimodality of the proposed metamaterial (see Section S4 of the Supporting Information for further details on these modes).

A parameter study is conducted to study the influence of the multimodal resonator design on the Dirac degeneracies presented in Figure 1c for the $\alpha$=0 case. Results from this parameter study (shown in Figure 2) illustrate how the respective frequencies $f_d$ of the Dirac degeneracies can be tailored by adjusting the frequency characteristics of the resonators. Figure 2a reveals that the Dirac frequencies have an inverse relationship with the resonator mass, which is adjusted using the mass height parameter $h_m$. By increasing the resonator mass, the Dirac frequencies for D1-D4 can be dramatically reduced, and the four degeneracies can be separated from extraneous high-frequency (>6 kHz) modes that would make it difficult to construct effective waveguides (see further details in Section S5 of the Supporting Information). By employing the local resonance mechanism to isolate the Dirac dispersions in this way, the proposed metamaterial unclutters the intrinsically dense band structure of 3D elastic metamaterials, which has often impeded low-frequency topological wave control in previous investigations. In contrast to the inverse relationship with the resonant mass, the Dirac frequencies exhibit a direct relationship to the stiffness of the resonator spring, which is modified through the ligament width $w_{l1}$ (Figure 2b). Notably, the $f_d$ for D3 is the most affected by increasing $w_{l1}$ from 1 mm to 3 mm (168% increase in $f_d$ for D3, compared to 37% for D1, 66% for D2, and 77% for D4). This large shift in D3 is due to the heightened sensitivity of the in-plane twisting mode to $w_{l1}$, compared to the other three sets of modes presented in Figure 1e. The cause of this greater sensitivity can be clarified by treating the spring ligaments as idealized cantilevered beams of length $L_l$. Under this assumption, the bending stiffness for the D1 mode is $S_{D1} = \frac{3w_{l1}h_l^3}{4L_l^3}$ [62], while for D3 $S_{D3} = \frac{3h_l w_{l1}^3}{4L_l^3}$ [62], explaining the dramatically enhanced influence of $w_{l1}$ on the frequency $f_d$ of the D3 degeneracy when compared with D1. These findings show how the unique stiffness mechanisms for each of the four resonant modes in the multimodal design can be exploited to tune the Dirac frequencies for D1, D2, D3, and D4 relative to each other. In addition, the results of this parameter study uncover how the multimodal resonance of the proposed metamaterial enables topological bandgaps at low frequencies (e.g., all $f_d$ < 4 kHz for $w_{l1}$ = 1 mm) that can be controlled without needing to alter the lattice constants $a$ or $h_o$, a significant advantage for volume-constrained applications.

### *Supercell Analysis for 2D Topological States*
To obtain 2D topological states, an eight-unit supercell is constructed that consists of four Type A unit cells connected to four Type B unit cells at a Type I interface (Figure 3a). Floquet periodic boundary conditions are applied along the $y-s$ and $z-s$ directions, while the left and right boundaries are left free. According to the bulk-boundary correspondence [1], [6], 2D topological states with displacement localized at the interface (i.e., interface states) are expected to emerge within the topological bandgaps, since the Type A and Type B lattices are topologically distinct. The band structure for the supercell is calculated along the surface Brillouin zone projected onto the $k_{y-s} - k_{z-s}$ plane (Figure 3b,c). A localization parameter $\Lambda_i = \iiint_{V_{interface}} d^2 \, dV / \iiint_{V_S} d^2 \, dV$ (where $V_{interface}$ is the volume of the two unit cells at the interface and $V_S$ is the total volume of the supercell) is defined to measure the confinement of the total displacement $d = \sqrt{|u|^2 + |v|^2 + |w|^2}$ at the interface for each eigenmode. Thus, interface modes have $\Lambda_i \approx 1$ (represented by the red bands in the band structure) and bulk modes have $\Lambda_i \approx 0$ (represented by the black bands in the band structure). As shown in Figure 3c, topological interface states are found in three distinct frequency ranges that align with the D2, D3, and D4 bandgaps. Representative mode shapes for each topological interface state are shown at the bottom of Figure 3c, illustrating rocking (found over a frequency range of 1.3 to 1.5 kHz), in-plane twisting (3.6 to 3.8 kHz), and in-plane translational (4.4 to 5.4 kHz) modes with displacement fields that match the unit cell resonant modes for D2, D3, and D4 displayed in Figure 1e. Similarly, a supercell is constructed from eight Type B unit cells, and the resulting band structure is shown in Figure 3d. The band structure for the Type B supercell is nearly identical to that of the supercell with the Type I interface. A localization parameter $\Lambda_b = \iiint_{V_{boundary}} d^2 \, dV / \iiint_{V_S} d^2 \, dV$ (where $V_{boundary}$ is the volume of the unit cells at the two supercell boundaries) is defined to measure the modal displacement contained at the boundaries of the Type B supercell. Rocking, in-plane twisting, and in-plane translational topological states are uncovered with displacements that are trapped at the left boundary of the supercell ($\Lambda_b \approx 1$, Figure 3d). Similar observations of topological boundary states that emerge due to a topological transition at the boundary of a lattice are reported in several previous works and predicted by the bulk-boundary correspondence [1], [6], [24], [63]–[66]. These topological boundary states emerge within the D2, D3, and D4 topological bandgaps and align with the frequency ranges for the interface states reported in Figure 3c. The findings gleaned from this supercell analysis reveal that the multimodal resonance of the proposed metamaterial facilitates the achievement of multiband topological interface and surface states.

Further inspection of the band structures in Figure 3c,d reveals that there are extraneous bulk modes that cross through the light gray D2, D3, and D4 bandgaps. While care must be taken to excite the D3 topological interface state without activating the nearby bulk modes, the superfluous bulk modes that cross through the D2 and D4 bandgaps near $\tilde{\Gamma}$ contain polarizations that preclude unwanted hybridization with the topological states (see more details in Section S6 of the Supporting Information). Another important feature of the supercell band structure is the sharp divot that is observed in the rocking (1.3 to 1.5 kHz) topological states for the $-\tilde{Z} - \tilde{\Gamma}$ and $\tilde{\Gamma} - \tilde{Z}$ directions. Since this steep slope (i.e., large group velocity) occurs near $\tilde{\Gamma}$ and the band is flat for the rest of the $-\tilde{Z} - \tilde{\Gamma} - \tilde{Z}$ wavenumber region, this indicates that the interconnecting rods undergo quasi-rigid body (or very long wavelength) motion that transmits wave energy along the $z$ direction (see Figure S8 of the Supporting Information). In contrast, the two in-plane topological states are flat (i.e., zero group velocity) and/or gapped for the $-\tilde{Z} - \tilde{\Gamma} - \tilde{Z}$ wavenumber region, and thus do not transmit wave energy along $z$. This dichotomy in $z$ direction transport behavior for the different topological states opens the door to polarization- and layer-dependent wave control functionality in full-scale 3D structures.

### *Topological Waveguides in Full-Scale 3D Metastructures*
Full-scale finite element simulations are conducted to investigate the elastic wave control capabilities of the metamaterial in finite 3D metastructures (a metastructure is defined in this paper as a mechanical structure created from the metamaterial that has finite boundary conditions). A 3D metastructure is constructed from an 8x8x6 tessellation of the metamaterial unit cell (Figure 4a). The four corners at the base of the metastructure are fixed and all other boundaries are left free. Two different elastic waveguides are created in the metastructure by specifying the distribution of Type A and Type B unit cells. The V-shaped waveguide (blue shading in the middle schematic in Figure 4a) is created from the topological boundary states, while the Z-shaped waveguide (blue shading in the right schematic in Figure 4a) is generated by connecting the topological boundary and interface states in series. A harmonic excitation is placed on all six metastructure layers (L1-L6) at one end of the waveguide. The excitation polarization and frequency are selected to activate each of the three unique topological states: out-of-plane at 1.3 kHz for the rocking state, in-plane at 3.6 kHz for the in-plane twisting state, and in-plane at 4.9 kHz for the in-plane translational state. The resulting steady-state displacement fields (Figure 4b) reveal a dynamic vibration response that is confined within the designated 2D waveguides and is polarized according to the corresponding topological state ($\Pi=0.6$ for the rocking waveguides and $\Pi\approx0$ for the in-plane waveguides). These topological waveguides exhibit unconventional polarization- and frequency-dependent behaviors. For example, if an out-of-plane excitation is applied in the frequency range for an in-plane topological state, the wave is attenuated by the topological bandgap and will not propagate beyond the input location (Figure 4c). Moreover, if the input excitation is provided on layers 1 and 4 (L1 and L4) of the metastructure, the in-plane topological states display layer-locked wave propagation (Figure 4d). The dynamic response is locked to L1 and L4 and there is no transfer of wave energy along the $z$ direction, which aligns with the supercell band structure for these two topological states. In the case of the rocking state, the waves propagate along all six layers for a two-layer input with a frequency of 1.3 kHz and undergo layer-locked transport if the two-layer input frequency is increased to 1.4 kHz (Figure 4d). This frequency-selective behavior is explained by examining the supercell band structure, which predicts that the waves will propagate in the $z$ direction for 1.28 to 1.33 kHz and exhibit more layer-locked behavior from 1.33 to 1.46 kHz due to the sharp divot that appears over $-\tilde{Z}-\tilde{\Gamma}-\tilde{Z}$ (Figure 3c,d, see more detail in Section S6 of the Supporting Information). These results illuminate how the multimodal topological states of the 3D metamaterial can be harnessed for polarization-, frequency-, and layer-dependent wave control in finite 3D metastructures. As shown in Section S7 of the Supporting Information, this multimodal mechanism can also be exploited in a 2D topological metamaterial, underscoring the generalizability of the proposed concept.

*Experimental Realization*
An experimental investigation is undertaken to validate the theoretical predictions and implement the proposed 3D metamaterial in a practical setting. The experimental testbed is a 3D metastructure made up of a 4x4x4 tessellation of the Type B metamaterial unit cell (Figure 5a, see Section S8 of the Supporting Information for a detailed description of the fabrication and experimental testing). A supercell analysis (for a four-unit supercell) reveals a topological boundary state with a rocking polarization that emerges in the frequency range of 1.3 to 1.5 kHz (Figure 5b). This boundary state is employed to design a V-shaped waveguide in the 3D metastructure, as indicated by the blue shading in the schematic located in the top right corner of Figure 5c. A periodic chirp vibration input with a bandwidth of 0 to 5 kHz is provided at the middle two layers (L2 and L3) of the metastructure where indicated in Figure 5c. The excitation is provided by a piezoelectric (lead zirconate titanate, or PZT) actuator pair that provides a harmonic bending displacement in the $z$ direction. A scanning laser Doppler vibrometer (SLDV, Polytec PSV-500) is used to acquire non-contact measurements for the out-of-plane velocity magnitude $v_{op}=|\dot{w}|$, which only requires the use of a single laser. The measurements are obtained for all four layers L1-L4 of the metastructure by guiding the laser in between the gaps in the layers closest to the vibrometer head. An out-of-plane velocity $v_{op}$ field taken at $f_{m1}$ = 1.3 kHz illustrates the successful confinement of the dynamic response within the designated 2D waveguide and closely aligns with finite element simulations (Figure 5c, an animation of the dynamic response on L1 is found in Supplemental Movie S1). The frequency response is gathered for a point inside the waveguide (Point A in Figure 5c) and a point outside the waveguide (Point B in Figure 5c) on each of the four layers. The frequency response displayed in Figure 5d reveals that there is a wide frequency range (indicated by gray shading) where the out-of-plane velocity $v_{op}$ taken at the point inside the waveguide (Point A, results represented by solid lines) is significantly higher than the velocity at the point outside the waveguide (Point B, results represented by dashed lines) for all four layers (see Section S9 of the Supporting Information for additional frequency response data). These results represent a clear advancement, in that this is the first time that full-field dynamic response information illustrating a 2D topological waveguide has been experimentally acquired for a 3D mechanical structure.

The in-plane twisting and in-plane translational topological states are also experimentally characterized in the 3D metastructure. An in-plane twisting state (shown at the bottom of Figure 6a) is used to construct a waveguide that follows the two planar boundaries identified by the red shading in Figure 6a. A piezo-stack actuator is utilized to apply an in-plane periodic chirp (with a bandwidth of 0 to 10 kHz) excitation to L4 of the structure. The in-plane velocity components $\dot{u}$ and $\dot{v}$ are measured using a 3D SLDV (Polytec PSV QTec 3D) and the total in-plane velocity magnitude is calculated as $v_{ip} = \sqrt{|\dot{u}|^2 + |\dot{v}|^2}$. The in-plane measurements require the use of three lasers, and thus the only surface accessible for measurement is L4, the layer closest to the 3D SLDV head. The velocity field measured at $f_{m2}$ = 3.7 kHz shows that the dynamic response is effectively confined into the waveguide by the topological twisting state and matches finite element simulations (Figure 6b, animations are found in Supplemental Movie S2 and S3). The frequency response measured for a point inside the waveguide and a point outside the waveguide (see the schematic in Figure 6a) reveals that waveguiding behavior occurs over the frequency range of 3.5 to 3.9 kHz (Figure 6c), which is enclosed within the bandwidth (3.5 to 3.7 kHz) of the in-plane twisting state. The same experimental approach is followed for the in-plane translational topological state. A waveguide is constructed to follow the blue surfaces in Figure 6d and experimental measurements illustrate successful waveguiding at $f_{m3}$ = 4.4 kHz (Figure 6e, animations are found in Supplemental Movie S4 and S5). Results displayed in Figure 6f indicate that the wave field is confined along the prescribed path over a frequency range of 4.3 to 5.0 kHz, which overlaps with the bandwidth (4.4 to 5.3 kHz) of the in-plane translational state reported in Figure 6d. The outcomes from this experimental study advance beyond the previous research on 3D topological metamaterials, which has overwhelmingly concentrated on the theoretical investigation of a single topological state [36]–[42], to attain experimental wave fields of topological waveguides with multiple (i.e., three: rocking, in-plane twisting, and in-plane translational) distinct polarizations and frequency bandwidths.

**Discussion**

The research presented in this paper advances the state of the art through the creation of a 3D topological metamaterial that exploits multimodal local resonance to enable low-frequency and multiband elastic wave manipulation. A band structure analysis reveals four topological states that emerge in distinct frequency regions that are associated with out-of-plane translational, rocking, in-plane twisting, and in-plane translational resonances. A parameter study is used to develop a deeper understanding of the connection between the frequency range for each topological state and the fundamental characteristics of the respective resonant modes (i.e., the mass and stiffness for each mode). This parametric analysis illuminates how the topological states can be attained over a broad bandwidth and at low frequencies (e.g., all below 6 kHz) by adjusting the mass and stiffness properties of the local resonators, without needing to increase the metamaterial volume. Finite element simulations of full-scale 3D metastructures uncover how the topological states can be employed to achieve 2D planar and layer-locked waveguides with complex frequency- and polarization-dependent behaviors. To complement the theoretical findings, experiments are conducted to validate the numerical predictions and establish a new benchmark for the experimental investigation of topological phenomena in 3D mechanical structures. The experimental advancements reported in this paper include the attainment of complete wave fields that unequivocally illustrate 2D topological waveguides and the measurement of multi-polarized wave control in a 3D structure. These outcomes may inspire future experimental research on wave control and topological physics in 3D mechanical systems. The exceptional capabilities and intuitive design methodology of the proposed 3D metamaterial would improve performance in engineering applications that involve low-frequency and multiband elastic wave control, such as vibration energy harvesters and isolators [10]–[13], [19], [53], [54]. Moreover, the rich 3D, multiband, frequency-dependent, and polarization-dependent features of the metamaterial could be utilized to construct information-dense phononic circuitry (i.e., elastic wave-based) for mechanical computers, wave filters, and on-chip communications [14]–[17], [20]–[22], [67]. For example, the proposed metamaterial could be used as the building block of 3D phononic circuits with multiple working channels that route waves in a layer-dependent fashion based on the polarization and frequency of external inputs.

**Materials and Methods**

Full details for the numerical simulations are described in Section S1 of the Supporting Information. The experimental methods, including both fabrication and testing, are comprehensively described in Section S8 of the Supporting Information.

**Acknowledgments**

The authors thank Osama Jameel and Polytec for their support with experimental measurements and the use of their PSV QTec 3D SLDV. The authors acknowledge the financial support of the Air Force Office of Scientific Research under Award No. FA9550-21-1-0032. P.D. also acknowledges financial support from the Rackham Merit Fellowship and Rackham Graduate Student Research Grant at the University of Michigan.

**Figures**

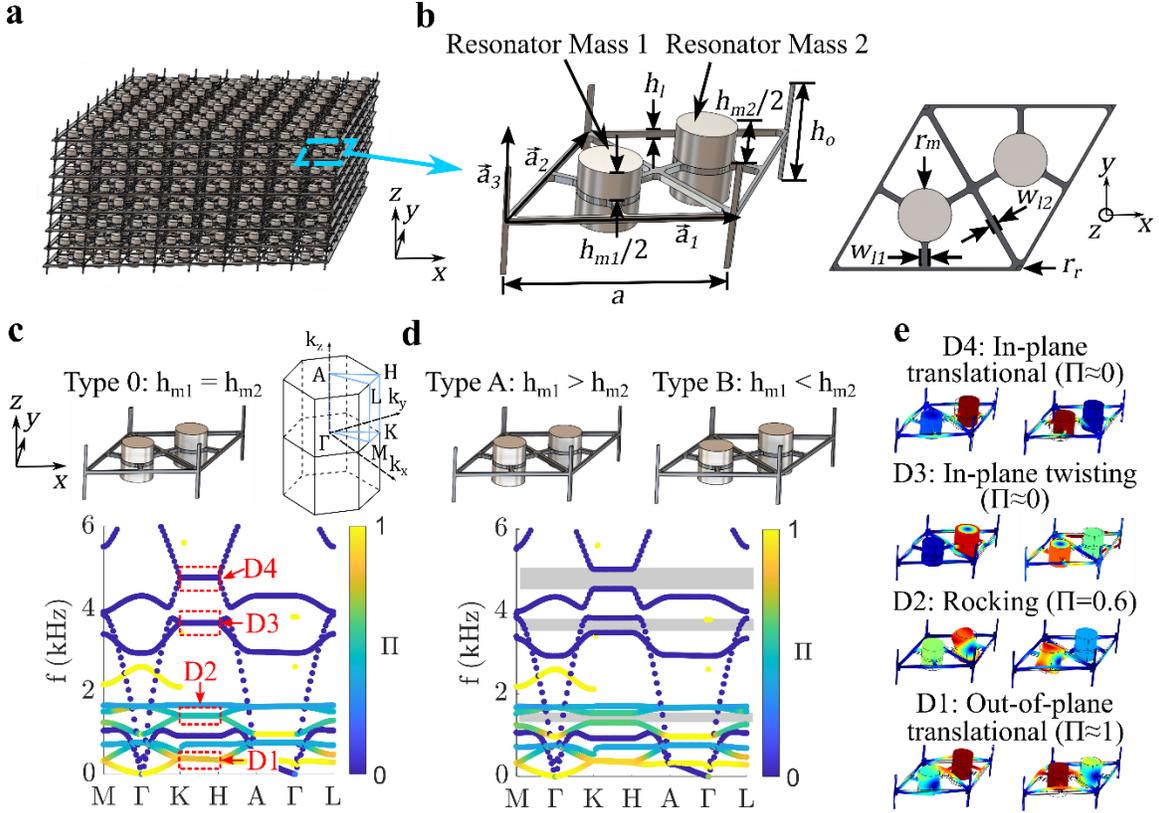

**Figure 1.** a) A schematic of the 3D topological metamaterial. b) Isometric and top views of the metamaterial unit cell. c) The band structure for the Type 0 lattice ($\alpha$ = 0). The four Dirac nodal line degeneracies are indicated by the dotted red boxes. The colorbar indicates the mode polarization quantified by the parameter $\Pi$. d) The band structure for the Type A/B ($\alpha$ = -0.11/0.11) lattices. The band structures for Type A and Type B lattices are identical and superimposed. The three topological bandgaps are shaded in gray. e) The mode shapes (taken along K-H) for the bands that border the four (D1, D2, D3, and D4) topological bandgaps in the $|\alpha|$=0.11 case, illustrating multimodal resonance.

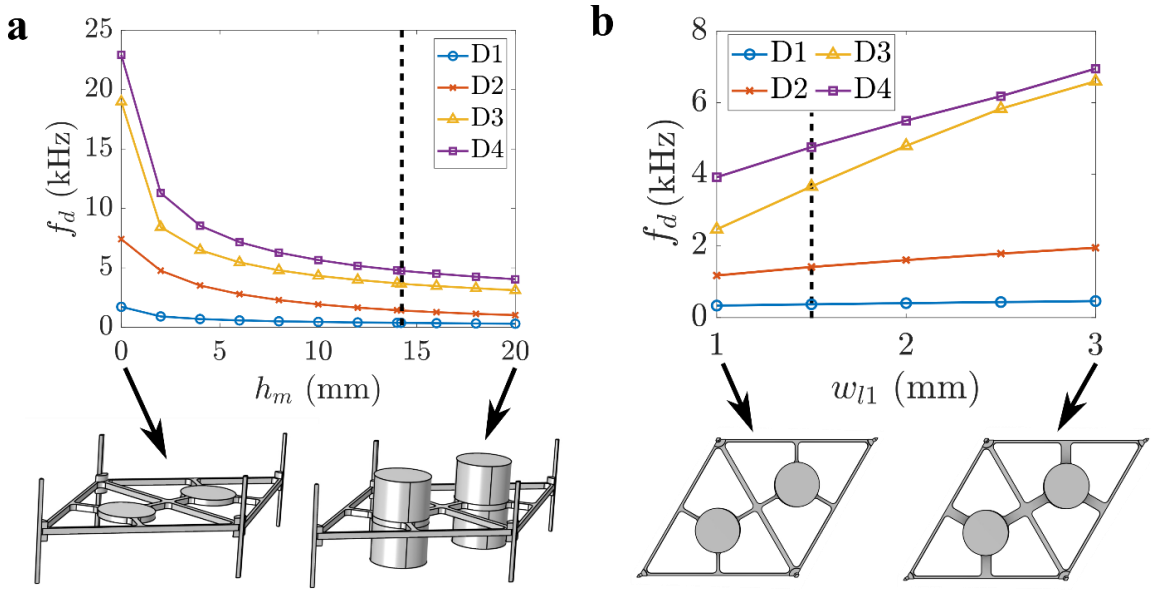

**Figure 2.** Parameter study illustrating the effect of the a) mass height $h_m$ and b) spring ligament width $w_{l1}$ on the Dirac nodal line frequency $f_d$, which is taken at the midpoint between K and H for each D1-D4. All presented values are for the Type 0 lattice configuration ($\alpha$ = 0). The insets show the unit cell geometries for the minimum and maximum specified values of $h_m$ and $w_{l1}$. The vertical dashed lines indicate the specified $h_m$ and $w_{l1}$ for all other presented results in this paper.

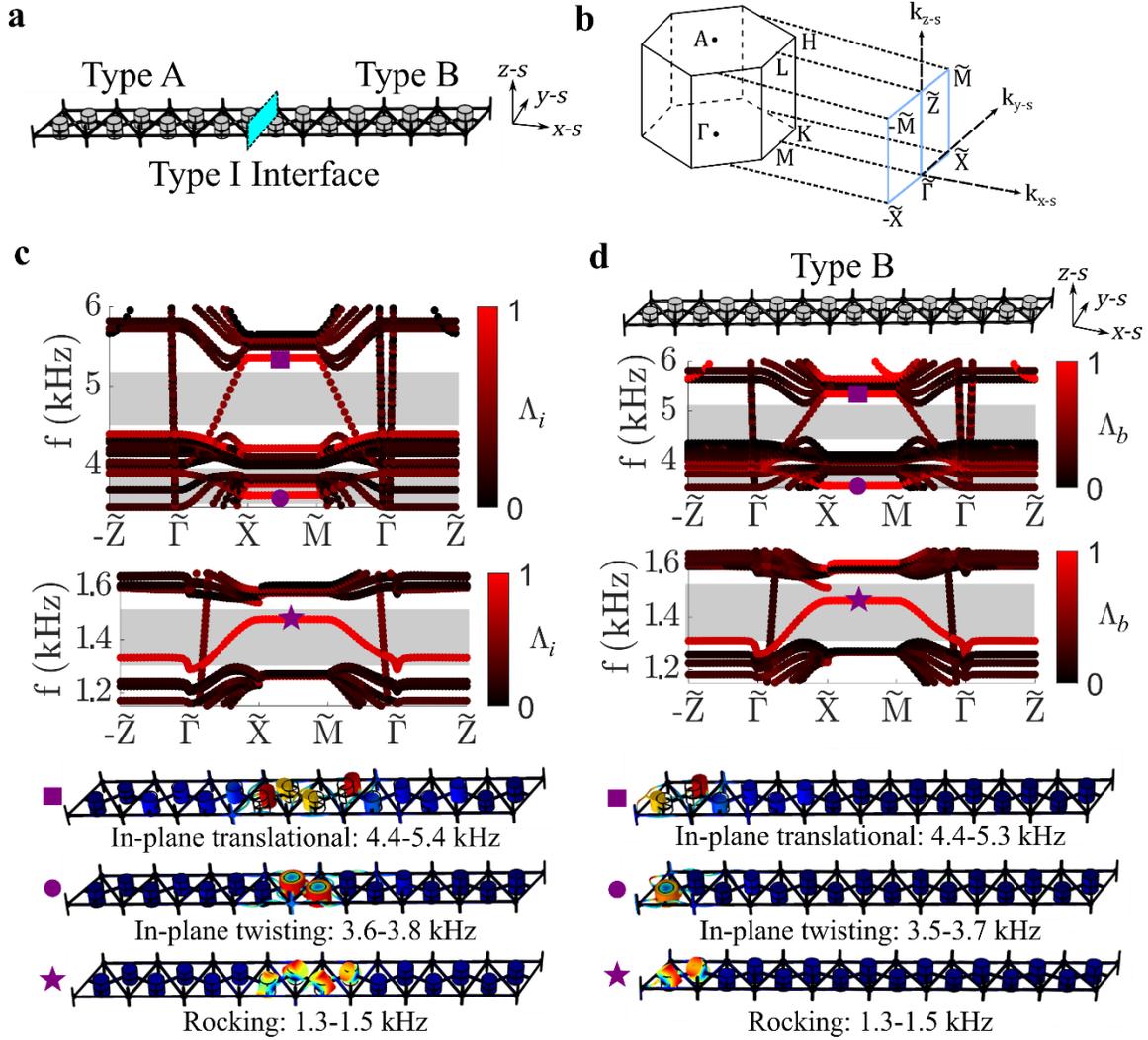

**Figure 3.** a) Schematic of an eight-unit supercell with a Type I interface indicated by the blue planar surface. b) The reciprocal space, with one-half of the surface Brillouin zone projected onto the $k_{y-s} - k_{z-s}$ plane outlined in light blue. c) The band diagram for the supercell presented in (a). The topological bandgaps are indicated by light gray shading. The red bands ($\Lambda_i \approx 1$) are interface modes and the black bands ($\Lambda_i \approx 0$) are bulk modes. Representative mode shapes for the rocking (purple star), in-plane twisting (purple circle), and in-plane translational (purple square) topological interface states are shown at the bottom. d) The schematic and band diagram for a supercell comprised of eight Type B unit cells. The red bands ($\Lambda_b \approx 1$) are boundary modes and the black bands ($\Lambda_b \approx 0$) are bulk modes. Representative mode shapes for the topological boundary states are shown at the bottom.

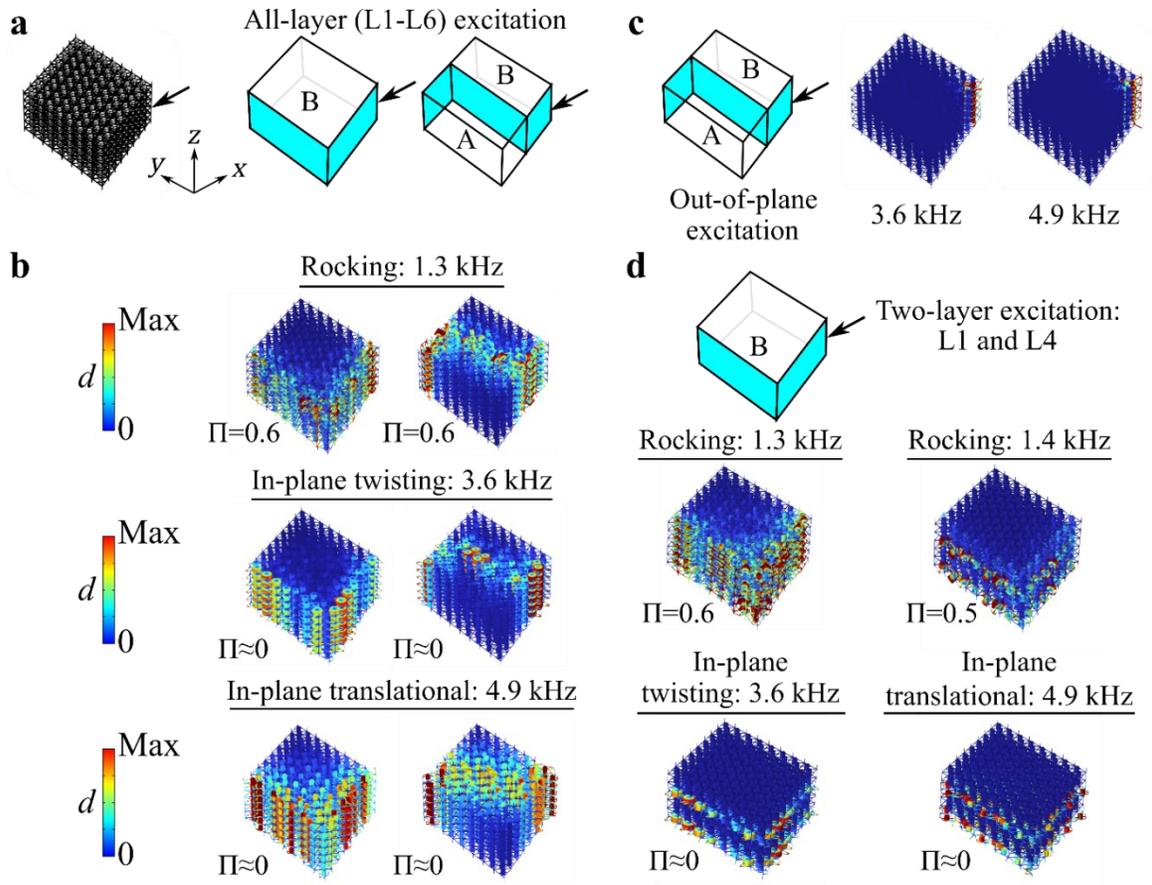

**Figure 4.** a) Schematic of a full-scale 3D metastructure constructed from an 8x8x6 pattern of the metamaterial unit cell (left) and illustrations of V-shaped (middle) and Z-shaped (right) waveguides. The distribution of Type A and Type B unit cells is denoted by the letters "A" and "B." b) Steady-state displacement fields illustrating waveguides for all-layer (L1-L6) input excitations of 1.3 kHz (rocking), 3.6 kHz (in-plane twisting), and 4.9 kHz (in-plane translational). c) Wave attenuation when an out-of-plane excitation is used in the frequency ranges of the in-plane topological states. d) Steady-state displacement fields illustrating layer-locked waveguiding for two-layer (L1 and L4) input excitations of 1.4 kHz (rocking), 3.6 kHz (in-plane twisting), and 4.9 kHz (in-plane translational). For a two-layer input of 1.3 kHz, the rocking state exhibits wave transmission across all six layers.

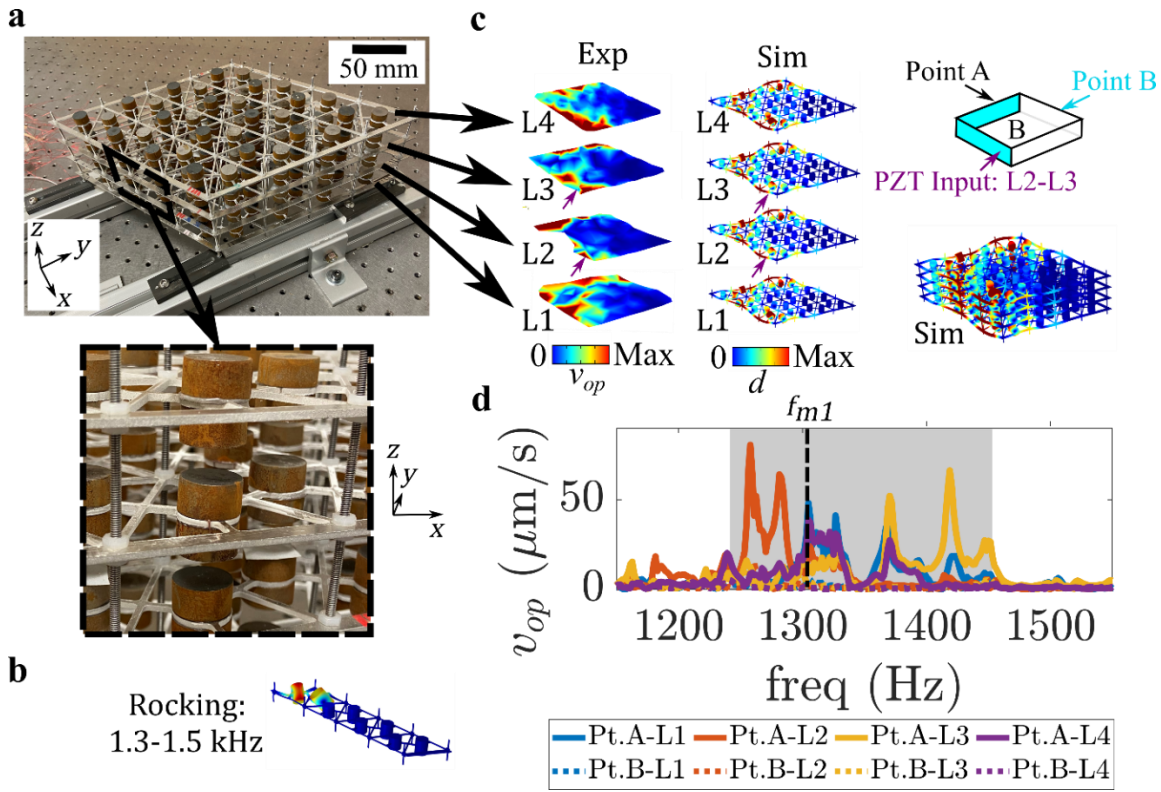

**Figure 5.** a) The experimental testbed with an inset showing detail for a unit cell. b) A topological boundary state with rocking polarization that is found in the band structure of a four-unit supercell. c) Experimentally measured out-of-plane velocity field (left) and finite element simulated displacement field (right) for the 3D metastructure obtained at $f_{m1}$ = 1.3 kHz. For clarity, both the full-scale and layer views of the simulated displacement field are shown. The schematic of the 3D metastructure testbed is given in the top right, where the blue shading represents the V-shaped waveguide. d) The experimentally measured out-of-plane velocity ($v_{op}$) for Point A (solid lines) and Point B (dashed lines) on each of the four layers. The frequency range for effective waveguiding is marked by the gray shading.

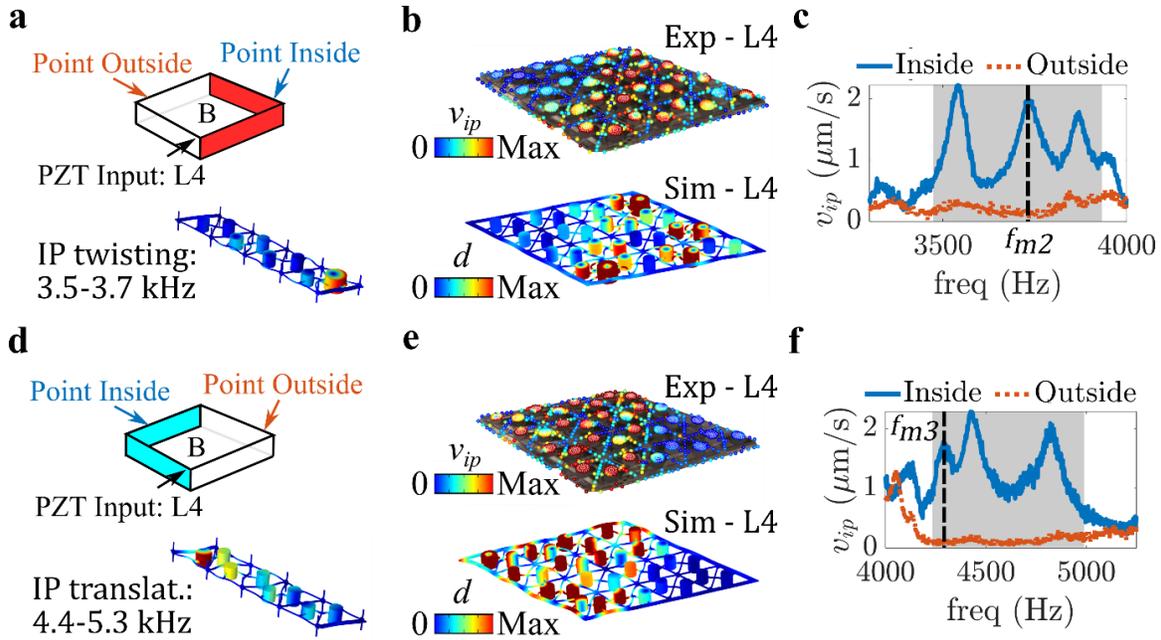

**Figure 6.** a) (top) A schematic of the 3D metastructure testbed where the red shading represents the path of a V-shaped waveguide. (bottom) A topological boundary state with an in-plane twisting polarization that is found in the band structure of a four-unit supercell. b) Experimentally measured in-plane velocity field and finite element simulated displacement field for L4 of the 3D metastructure obtained at $f_{m2}$ = 3.7 kHz. c) The experimentally measured in-plane velocity ($v_{ip}$) for a Point Inside and a Point Outside the waveguide on L4. The frequency range for effective waveguiding is marked by the gray shading. d) (top) A schematic of the 3D metastructure testbed where the blue shading represents the path of a V-shaped waveguide. (bottom) A topological boundary state with an in-plane translational polarization that is found in the band structure of a four-unit supercell. e) Experimentally measured in-plane velocity field and finite element simulated displacement field for L4 of the 3D metastructure obtained at $f_{m3}$ = 4.4 kHz. f) The experimentally measured in-plane velocity ($v_{ip}$) for a Point Inside and a Point Outside the waveguide on L4. The frequency range for effective waveguiding is marked by the gray shading. See Section S9 of the Supporting Information for the transmission ratio plots that accompany c) and f).

18